# Enhancement of strong-field multiple ionization in the vicinity of the conical intersection in 1,3-cyclohexadiene ring opening

Vladimir S. Petrovic[1,2,*], Sebastian Schorb[3], Jaehee Kim[1,2], James White[2,4], James P. Cryan[1,2,3], J. Michael Glownia[2,3,4], Lucas Zipp[1,2,3], Douglas Broege[2,3,4], Shungo Miyabe[2,5], Hongli Tao[2,5], Todd Martinez[2,3,5], Philip H. Bucksbaum[1,2,3,4]

[1]Stanford University, Department of Physics, Stanford, CA 94305
[2]The Stanford PULSE Institute, Menlo Park, CA 94025
[3]SLAC National Accelerator Laboratory, Menlo Park, CA 94025
[4]Stanford University, Department of Applied Physics, Stanford, CA 94305
[5]Stanford University, Department of Chemistry, Stanford, CA 94305

Nonradiative energy dissipation in electronically excited polyatomic molecules proceeds through conical intersections, loci of degeneracy between electronic states. We observe a marked enhancement of laser-induced double ionization in the vicinity of a conical intersection during a non-radiative transition. We measured double ionization by detecting the kinetic energy of ions released by laser-induced strong-field fragmentation during the ring-opening transition between 1,3-cyclohexadiene and 1,3,5-hexatriene. The enhancement of the double ionization correlates with the conical intersection between the HOMO and LUMO orbitals.

## I. INTRODUCTION

Nonradiative relaxation of electronic excitation in polyatomic molecules proceeds via conical intersections [1], loci of degeneracy between different electronic states. The Born-Oppenheimer approximation, which is suitable for describing a molecule in equilibrium geometry, breaks down in the vicinity of conical intersections. When that occurs, the equations describing motion of electrons and nuclei become coupled, leading to difficulties in computing the molecular dynamics. Due to the prevalence of conical intersections, there has been considerable effort to develop suitable theoretical frameworks and new experimental techniques to probe them [1]. In this article we report the observation of enhanced multiple ionization in the vicinity of conical intersections, and discuss how strong-field ionization could probe the properties of non-radiative molecular transitions.

Intense laser fields can initiate molecular fragmentation through multiple ionization followed by a repulsion of the remaining positive charges [2]. The ability of the molecular fragmentation to report on a chemical transformation in time-resolved experiments depends on the difference in the fragmentation patterns between the starting material and the transient species or product. Such time-resolved fragmentation experiments have been used to follow wavepacket motion [2] and study unimolecular reactions [3, 4].

---

[*] petrovic@stanford.edu

Photoinitiated ring opening of 1,3-cyclohexadiene (CHD) to form 1,3,5-hexatriene (HT) is a paradigmatic example of a unimolecular reaction that occurs through a conical intersection between two electronic states ($S_1$ and $S_0$). The molecule has been studied extensively [3-13] (for a recent review see [14] and the references therein), and is suitable for testing new experimental techniques. Ion fragment time-of-flight (TOF) mass spectra of the parent molecule CHD and the product HT differ significantly when short intense 800 nm radiation is used for fragmentation [3, 4, 15]. The HT isomer is characterized by a more violent fragmentation that leads to an increase of smaller-mass fragments, particularly $H^+$, compared to that of CHD. This qualitative difference between the CHD and HT fragmentation TOF mass spectra has been used in time-resolved experiments to monitor the CHD ring opening [3, 4].

The experiment reported here can distinguish fragmentation channels originating from different charge states $CHD^{n+}$ by measuring the kinetic energy of the fragment groups. Past measurements only monitored the evolution of the ion-time-of-flight (TOF) spectra following photoexcitation. We observed an enhancement in the double ionization, which correlates with the passage of the wavepacket through the conical intersection that occurs due to degeneracy between the HOMO and LUMO orbitals.

## II.     EXPERIMENT

To investigate the sensitivity of strong-field multiple ionization to conical intersections, we initiated the isomerization of CHD by a short UV pulse (266 nm, ~120 fs, <10µJ, produced by two-stage nonlinear mixing of 800 nm pulses). The relatively long duration of this excitation pulse avoids impulsive vibrational excitation that would add complexity to the interpretation of the results. The excitation and fragmentation lasers were collinear and orthogonally polarized. The laser beams intersected an effusive molecular beam at 90 degrees.

The ultraviolet excitation pulse launches a wavepacket on a spectroscopically bright $S_2$ potential energy surface that rapidly crosses onto the $S_1$ surface and accelerates to the $S_1/S_0$ conical intersection. We monitored the wavepacket crossing the $S_1/S_0$ conical intersection using time-resolved fragmentation initiated by an intense IR field (800 nm, ~80 fs, 100 µJ). Ions were collected by a velocity map imaging (VMI) detector (already described in [16]) with the following voltages applied to the repeller plate, extractor plate and flight tube, respectively: 1960 V, 1600 V, 1000 V. A portion of the electric field applied to the front of the micro-channel plate detector was pulsed to create a temporal gate in order to separate different fragments based on their travel times. The detector axis was perpendicular to the plane defined by the laser beams and the molecular propagation axis. The images were recorded by a Thorlabs DC210 camera (640x480 pixels). Kinetic energies of different ion fragment groups, separated by their travel times, were measured in steps of 30 fs over ~1.5 ps. We adjusted the gate timing to measure the fragmentation groups $H^+/H_2^+$, $CH_n^+$, $C_2H_n^+$, $C_3H_n^+$, $C_4H_n^+$, $C_5H_n^+$ and $C_6H_n^+$. We did not attempt to separate individual peaks within each group. Based on the fluctuations in the ion count in the raw signal, we estimate the relative standard error of our measurement to be on the order of 4%.

After finding the centroids of the hits on the detector and centering the averaged velocity map images we inverted them using the onion peeling algorithm [17]. This procedure was performed for all of the ion fragment groups, except for the $C_6H_n^+$ group, where the ion count was too large to permit the centroid-picking algorithm to work and instead raw images were averaged, centered, and inverted. The kinetic energy axis was calibrated based on a SIMION calculation for the experimental geometry. The accuracy of the measurement of lower kinetic energy fragments is limited by lower efficiency of our microchannel plates close to the center of the detector, as well as imperfections in the VMI focusing, image centering, and the inversion algorithm. The inability to perform the ion-hit centroid finding in the case of the parent ion due to a high-count rate contributes to the error of the kinetic energy measurement in that case.

### III. RESULTS AND DISCUSSION

Figure 1 shows the ion kinetic energy dependence on the time delay between the UV pump pulse and the IR fragmentation probe pulse, plotted separately for different ion fragment groups. Two sets of fragments, differentiated by their kinetic energy, are visible in $C_2H_n^+$, $C_3H_n^+$, and $C_4H_n^+$ ion groups. We refer to these sets in further discussion as 'low kinetic energy' (~1eV) and 'high kinetic energy' (2 - 12 eV) sets. In other ion groups only one of the two sets is apparent. This is the case in $H^+$ and $CH_n^+$ groups, which display only the high kinetic energy fragment sets, and $C_6H_n^+$ group where only the low kinetic energy fragment set is observed. In some cases a partial overlap prevents a better separation of different kinetic energy sets. This is the case in $C_5H_n^+$, where the peak of the distribution is at lower kinetic energy, but a long tail persists at higher kinetic energies.

Separation of ion fragments into the high and low kinetic energy groups allows us to infer the ionic charge of $CHD^{n+}$ at the onset of the fragmentation. We estimate that, depending on the location of the charges at the onset of fragmentation, the kinetic energy divided between two fragments in the fragmentation of $CHD^{2+}$ ranges from 5-13 eV (respectively, for limiting cases of two positive charges located on two farthermost carbons 2.9 Å apart, or neighboring C and H atoms separated by 1.1 Å). The kinetic energies of the high kinetic energy fragment groups in $H^+$, $CH_n^+$, $C_2H_n^+$, $C_3H_n^+$, and $C_4H_n^+$ suggest that these ions are formed by fragmentation of $CHD^{2+}$ (excluding a very weak peak in $CH_n^+$ at 8.2 eV, not visible in the color scheme of the Fig. 1, likely originating from $CHD^{3+}$). In fragmentation of a doubly charged ion the heavier fragment is released with less kinetic energy. This agrees with a decrease in the separation between the low and high kinetic energy peaks. The low kinetic energy in fragment sets visible in the $C_2H_n^+$, $C_3H_n^+$, $C_4H_n^+$, $C_5H_n^+$, and $C_6H_n^+$ groups suggest that these ion fragments resulted from a dissociation of excited CHD$^+$, which can be formed by multiple single- and two-color pathways.

The time dependence of the integrated signals for higher and lower kinetic energy fragment groups are plotted in Fig. 2. Three patterns of behavior are observed: lower kinetic energy fragments that peak at zero time delay (in C$_2$H$_n^+$, C$_3$H$_n^+$, and C$_4$H$_n^+$, C$_5$H$_n^+$, C$_6$H$_n$ groups), higher kinetic energy fragments that peak 150 fs after zero delay (in H$^+$, CH$_n^+$, C$_2$H$_n^+$, C$_3$H$_n^+$, C$_4$H$_n^+$ groups), and higher kinetic energy fragments that peak at zero delay (part of CH$_n^+$ group, possibly tail of the C$_5$H$_n^+$ group). Based on their different time dependence and kinetic energy, we propose that three time-dependent processes contribute appreciably to the signal in our experiment: one originating from decomposition of CHD$^+$; and two from the decomposition of CHD$^{2+}$. These three processes are: multi-photon single ionization; IR-ionization of the UV-formed CHD$^+$; and IR double ionization of the UV-excited neutral CHD/HT system. We will now discuss the details of the time dependence for the rates of these three processes, with particular attention on the IR double-ionization of the UV-excited neutral CHD/HT system. We are especially interested in the highest kinetic energy fragments and we concentrate the discussion on the primary fragmentation processes.

The first process that our analysis identifies is multi-photon single ionization of CHD. This process is responsible for the low energy ionic fragments formed by predissociation of CHD$^+$. Although the singly charged parent ion with sufficient excess energy to fragment can be formed by either absorption of two UV photons (IP$_{CHD}$ = 8.25 eV [14]), or by combination of the UV and IR photons (the 800 and 266 nm photons correspond to 1.55 eV and 4.65 eV, respectively), only the two-color signal displays a dependence on the time delay. The single-color pathway contributes to the time-independent baseline signal not discussed here, while the signal corresponding to two-color pathways is expected to peak at time delays when both pulses are present, as observed.

The second process that contributes to our signal is the sequential IR-ionization of UV-formed CHD$^+$. This agrees with our earlier results reported in [4], where we observed a significant contribution from the UV-formed CHD$^+$. The kinetic energy of the fragments produced by sequential two-color ionization corresponds to that of CHD$^{2+}$ fragmentation, and the prompt appearance of these fragments supports our assignment of this process. Large portions of the signal in the C$_5$H$_n^+$ and CH$_n^+$ channel result from this process.

A third process displays a maximum ~150 fs after the photoexcitation. The delayed peaks occur only in a subset of fragments originating from CHD$^{2+}$, and therefore indicate enhanced direct strong-field double ionization by the IR pulse of the UV-excited neutral CHD/HT system. This process is responsible for the higher kinetic energy fragments observed in the H$^+$, C$_2$H$_n^+$, C$_3$H$_n^+$, C$_4$H$_n^+$ ion groups, and to some extent signal in the C$_5$H$_n^+$ and CH$_n^+$ ion groups. An experiment reported in [16] finds that kinetic energy release, averaged over *all* of the fragments, goes through a delayed maximum before reaching the value corresponding to the CHD and HT mixture at late times [16]. That maximum is absent when x-rays are used to initiate the fragmentation [16], suggesting that a strong-field process is responsible for the delayed enhancement. This supports our assignment of the third process.

We propose that the two paths for CHD$^{2+}$ fragmentation discussed in the second process (sequential two-color double ionization) and third process (direct single-color double ionization) could come from different intermediates. These different intermediates are the singly charged CHD$^+$ and the doubly charged CHD$^{2+}$, respectively. Sequential two-color double ionization proceeds through CHD$^+$, which is predicted to relax by a hydrogen loss [18] to form a benzenium ion $C_6H_7^+$. On the other hand, CHD$^{2+}$ formed directly by single-color strong-field double ionization stabilizes by proton migration [18]. As a consequence, Coulomb repulsion produces fragments from $C_6H_7^{2+}$ in the sequential two-color process, and from $C_6H_8^{2+}$ in the direct process. The positive charge in $C_6H_7^+$ is distributed over five of the six carbon atoms, so the finding that $C_5H_n^+$ and $CH_n^+$ are produced in the high-kinetic energy data supports our attribution of this process to sequential two-color double ionization. Ion-ion coincident fragmentation measurements could provide a more detailed account of the discussed processes.

Figure 2 indicates that double ionization of the UV-excited CHD/HT system peaks at about 150 fs time delay. Previous experiments and theoretical models that have investigated the $S_0/S_1$ conical intersection find that the wavepacket launched by the UV pulse passes through a conical intersection at about this same time, approximately 140 fs after the photoexcitation [3, 14, 19]. This may indicate that when the $S_1$ and $S_0$ states become nearly degenerate, the double ionization rate is strongly enhanced. In that case, enhanced double ionization would lead to an overall increase in fragmentation, manifested as an increase in lower-mass fragments and an increase in the fragment kinetic energy. Indeed, previous work by ourselves and others on CHD isomerization [4, 13, 15, 20] also reports that proton and low mass fragment ejection goes through a delayed maximum, before reaching the value corresponding to the CHD and HT mixture at late times. The persistent high kinetic energy of lower-mass fragments suggests that the double ionization rate remains higher after the maximum compared to that of CHD. This is expected from the resonance in the HT$^+$ ion for absorption of 800 nm [21].

We propose that the observed enhancement in the double ionization could be related to the degeneracy between the HOMO and LUMO orbitals that is reached as the molecular geometry changes during the isomerization. Ab initio modeling of this process requires a treatment of the interaction of a strong field with a two-electron system that is strongly coupled to a change in the nuclear geometry. Such calculation is currently beyond reach even for smaller molecules than CHD. In absence of an ab initio model, we speculate that the observed enhancement in the double ionization could be related to the phenomenon of enhanced ionization previously reported in strong-field ionization of diatomic molecules [22, 23] (see Supplemental Material for details [24]). The new experimental observation that the cross section for the double ionization displays a delayed maximum upon the photoexcitation of CHD will be important in testing future models of strong-field ionization of photoexcited polyatomic molecules.

A similar enhancement in proton ejection has been reported in strong-field fragmentation of acetylene in the ground state [25]. The model discussed in that work proposes a mechanism in which a near-degeneracy of the HOMO and HOMO-1 orbitals, which occurs in the limit of strong-field extended C-H bonds, enhances the multiple ionization that results in proton detachment. In the acetylene example [25] the degeneracy between the two relevant orbitals, HOMO and HOMO-1, comes from the C-H bond elongation by the intense laser field. In CHD the degeneracy of the HOMO and LUMO orbitals for a particular geometry exists even in zero field, but it is expected that the strong-field parameters have an effect on the characteristics of the delayed enhancement [26]. Indeed, in the experiment reported here the delayed peak occurs 150 fs after the cross correlations, while earlier experiments report delays of up to 200 fs. We plan to further explore the effect of the strong field parameters on the orbital degeneracy and its potential for control over nonradiative relaxation.

Both our work and the account reported in Ref. [25] support the findings of Roither et al. who find that proton ejection is a dominant fragmentation channel in strong-field ionization of small hydrocarbons [27]. The enhanced multiple ionization, observed to occur in the vicinity of the $S_1/S_0$ conical intersection in CHD, leads to an increase in fragmentation that is manifested as a rise in the proton abundance. Protons, as the lightest fragments, depart with highest kinetic energy and show a particularly marked change upon the onset of the double ionization. In the experiment discussed here, we report that some of the other ions follow the same pattern as $H^+$.

## IV.  CONCLUSIONS

We observe an enhancement in strong-field multiple ionization that correlates with a valence orbital degeneracy during the ring opening of CHD. In particular, time-resolved measurements of kinetic energy release for different fragments reveal rates for competing processes in strong-field ionization. The time evolution of these rates provides information about transient structural changes in the molecule. In CHD the observed increase in the double ionization rate during the passage of the molecule through the $S_1/S_0$ conical intersection geometry leads to an increase in the release of energetic fragments during strong-field fragmentation. Specifically, the proton channel shows a notable change upon the onset of double ionization. Our work supports the findings reported in [25, 27] that the release of the energetic protons is a general feature of the strong-field fragmentation of small hydrocarbons.

Theoretical modeling of the observed enhancement in the double ionization rate requires models that can treat strong-field two-electron dynamic processes. Such *ab initio* models are difficult even for atoms. The experimental findings that are reported here will be important in testing future models of strong-field ionization of photoexcited polyatomic molecules. Future experiments that benefit from higher time resolution and variation of the strong-field laser parameters will help understand the mechanism of the enhanced strong-field multiple ionization in the vicinity of conical intersections. As conical intersections constitute the primary channel for nonradiative relaxation in polyatomic molecules [1], we expect that the method described here will find a widespread use in investigations of molecular structure and dynamics.

V. S. P., J. K., J. L. W. and P. H. B. were supported by the National Science Foundation, grant number PHY-0969322. S. S. received his support through the Linac Coherent Light Source at SLAC National Accelerator Laboratory, Office of Basic Energy Sciences, U.S. Department of Energy. J. P. C., J. M. G., L. Z., D. B., S. M., H. T. and T. M. were supported by the AMOS program within the Chemical Sciences, Geosciences, and Biosciences Division of the Office of Basic Energy Sciences, Office of Science, U.S. Department of Energy, under Contract DE-AC02-76SF00515. We thank Christoph Bostedt, John Bozek, and Markus Gühr for assistance with the VMI detector.

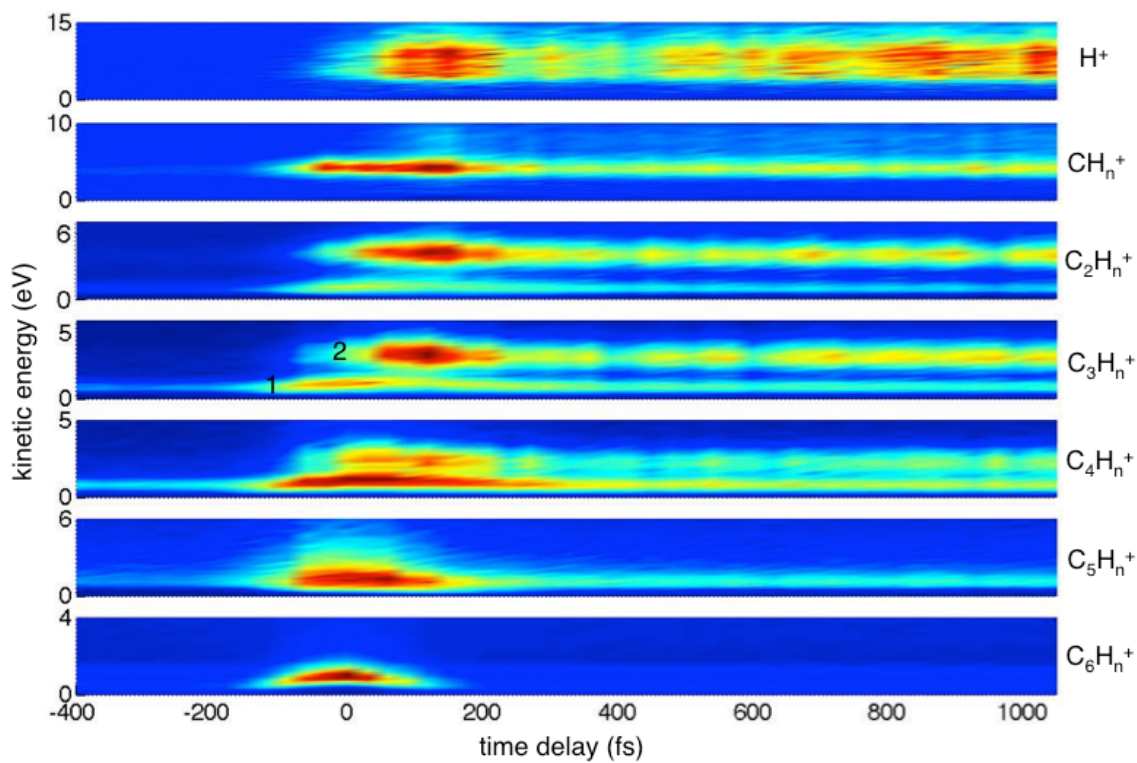

Fig. 1 Dependence of the kinetic energy of $H^+$, $CH_n^+$, $C_2H_n^+$, $C_3H_n^+$, $C_4H_n^+$, $C_5H_n^+$ and $C_6H_n^+$ fragment groups on the time delay between the UV and IR pulses (blue represents low count, red high count; color scale differs between plots). Low and high kinetic energy sets are labeled 1 and 2, respectively, in the case of $C_3H_n^+$.

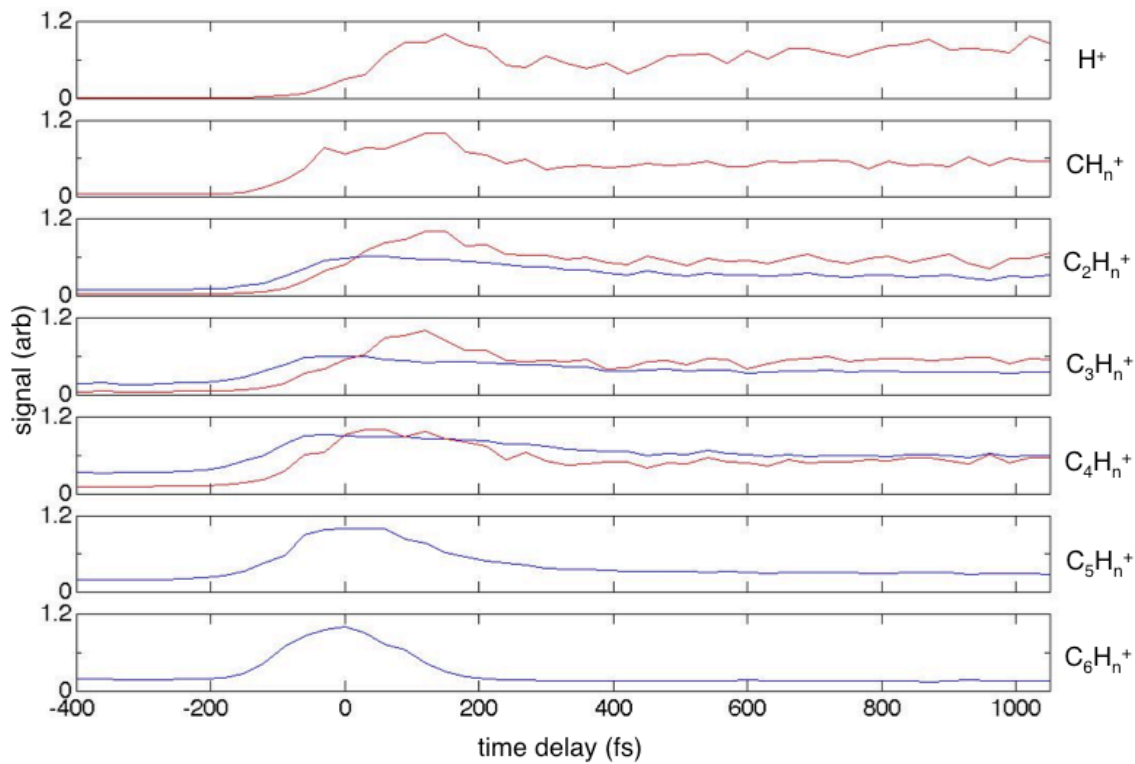

Fig. 2 Traces of integrated signal for 'low' (blue) and 'high' kinetic energy (red) sets in $H^+$, $CH_n^+$, $C_2H_n^+$, $C_3H_n^+$, $C_4H_n^+$, $C_5H_n^+$ and $C_6H_n^+$ fragment groups.

[1] W. Domcke, D. Yarkony, and H. Köppel, Conical Intersections: Electronic Structure, Dynamics & Spectroscopy, World Scientific, 2004.
[2] C. Ellert, H. Stapelfeldt, E. Constant, H. Sakai, J. Wright, D. Rayner, and P. Corkum, Philosophical Transactions of the Royal Society A: Mathematical, Physical and Engineering Sciences 356 (1998) 329.
[3] K. Kosma, S. Trushin, W. Fuß, and W. Schmid, Physical Chemistry Chemical Physics 11 (2009) 172.
[4] J. L. White, J. Kim, V. S. Petrović, and P. H. Bucksbaum, The Journal of Chemical Physics 136 (2012) 054303.
[5] M. Merchan, L. Serrano-Andres, L. Slater, B. Roos, R. McDiarmid, and X. Xing, in J. Phys. Chem. A, Vol. 103, 1999, p. 5468.
[6] S. Pullen, N. Anderson, L. II, and R. Sension, in J. Chem. Phys., Vol. 108, 1998, p. 556.
[7] M. Kotur, T. Weinacht, B. Pearson, and S. Matsika, The Journal of Chemical Physics 130 (2009) 134311.
[8] H. Ihee, V. Lobastov, U. Gomez, B. Goodson, R. Srinivasan, C. Ruan, and A. Zewail, Science 291 (2001) 458.
[9] C. Ruan, V. Lobastov, R. Srinivasan, B. Goodson, H. Ihee, and A. Zewail, Proceedings of the National Academy of Sciences 98 (2001) 7117.
[10] A. Hofmann, L. Kurtz, and R. de Vivie-Riedle, Applied Physics B: Lasers and Optics 71 (2000) 391.

[11] P. Celani, F. Bernardi, M. Robb, and M. Olivucci, in J. Phys. Chem, Vol. 100, 1996, p. 19364.
[12] H. Tamura, S. Nanbu, T. Ishida, and H. Nakamura, The Journal of Chemical Physics 124 (2006) 084313.
[13] W. Fuss, W. Schmid, and S. Trushin, The Journal of Chemical Physics 112 (2000) 8347.
[14] S. Deb and P. Weber, Annual Review of Physical Chemistry 62 (2011) 19.
[15] S. Trushin, W. Fuß, T. Schikarski, W. Schmid, and K. Kompa, The Journal of Chemical Physics 106 (1997) 9386.
[16] V. Petrović, M. Siano, J. White, N. Berrah, C. Bostedt, J. Bozek, D. Broege, M. Chalfin, R. Coffee, J. Cryan, L. Fang, J. Farrell, L. Frasinski, J. Glownia, M. Gühr, M. Hoener, D. Holland, J. Kim, J. Marangos, T. Martinez, B. McFarland, R. Minns, S. Miyabe, S. Schorb, R. Sension, L. Spector, R. Squibb, H. Tao, J. Underwood, and P. Bucksbaum, Physical Review Letters 108 (2012) 253006.
[17] G. Roberts, J. Nixon, J. Lecointre, E. Wrede, and J. Verlet, Review of Scientific Instruments 80 (2009) 053104.
[18] T. S. Zyubina, A. M. Mebel, M. Hayashi, and S. H. Lin, Physical Chemistry Chemical Physics 10 (2008) 2321.
[19] H. Tao, First Principles Molecular Dynamics and Control of Photochemical Reactions, Ph.D. Thesis, Stanford University 2011
[20] N. Kuthirummal, F. Rudakov, C. Evans, and P. Weber, The Journal of Chemical Physics 125 (2006) 133307.
[21] H. Harada, S. Shimizu, T. Yatsuhashi, S. Sakabe, Y. Izawa, and N. Nakashima, Chemical Physics Letters 342 (2001) 563.
[22] T. Zuo and A. D. Bandrauk, Physical Review A 52 (1995) R2511.
[23] T. Seideman, M. Y. Ivanov, and P. B. Corkum, Physical Review Letters 75 (1995) 2819.
[24] See Supplementary Material Document No.______ for [brief description]. For information on Supplementary Material.
[25] E. Lötstedt, T. Kato, and K. Yamanouchi, Physical Review A 85 (2012)
[26] J. Kim, H. Tao, J. L. White, V. S. Petrović, T. J. Martinez, and P. H. Bucksbaum, The Journal of Physical Chemistry A 116 (2011) 2758.
[27] S. Roither, X. Xie, D. Kartashov, L. Zhang, M. Schöffler, H. Xu, A. Iwasaki, T. Okino, K. Yamanouchi, A. Baltuska, and M. Kitzler, Physical Review Letters 106 (2011) 163001.

Supplemental Material

The observed enhancement of multiple ionization in CHD could potentially be related to the phenomenon of enhanced ionization previously reported in strong-field ionization of diatomic molecules [22, 23]. A pair of charge resonant states [24] coupled by the field can form an asymmetric distribution of the charge around the two nuclei, which can enhance ionization. The stretching of the internuclear separation by the strong field leads to localization of the bonding electron and enhanced tunnel ionization [22, 25]. It is possible that a similar situation arises in the enhanced ionization of the CHD. In the case of CHD the distortion of the molecular geometry from the equilibrium occurs upon UV excitation. Although in the case of the CHD the distortion of the molecular geometry is not as simple as extension of one bond, for the purpose of examining the effect the geometry change can have on the electron localization and consequently enhanced ionization, we will consider primarily the increase in the bond length between the two C atoms between which the bond breaks during the ring opening.

The extension of this idea to polyatomic molecules employs the concept of broken symmetry: When the molecular geometry changes following UV excitation, the original approximate $C_{2v}$ symmetry is broken, and the dipole coupling between $S_1$ and $S_0$ may therefore increase (see Fig. S1). Although a full analysis of this is beyond the scope of this paper, we have made a preliminary calculation. The transition dipole moment in Fig. S1 was calculated using the ab initio multiple spawning (AIMS) method, implemented over some 70 runs [18]. The trajectories were population weighted and each run contributed equally. At the conical intersection $S_1$ and $S_0$ are degenerate, and the coupling by the field is strongest. If the two C atoms between which the bond breaks are oriented along the field, this coupling can result in an asymmetric distribution of the charge around the uphill and downhill carbons. When the ring starts to open, the increase in the internuclear distance between the two carbons in this proposed mechanism results in a localization of the charge in presence of the strong field on one of the carbons. This localization enables tunneling that enhances the ionization. It is possible that the distortion of other bond lengths and angles contributes to the localization. A calculation of geometry of neutral and cation carried out with SA-2-CAS(6/6)-MSPT2/6-31G* and SA-2-CAS(5/6)-MSPT2/6-31G* methods, respectively, finds a conical intersection in the CHD cation in the vicinity of the geometry for the conical intersection in the neutral (see Fig. S2). A small distortion between these two geometries can easily occur in the strong field. This finding supports a similar mechanism for removal of the second electron.

It is also interesting to consider the possibility of two electrons being ejected simultaneously in the presence of the strong field. When the initial and the final state in the ionization are calculated in its own self-consisted framework [26] there will be non-zero overlaps for double ionization, which only increase in presence of electron correlation. In that case the term in the Hamiltonian for ejection of electrons $e_1$ and $e_2$ from orbitals $\phi_1$ and $\phi_2$ into the continua $\varepsilon_1$ and $\varepsilon_2$ is:

$$\langle\phi_1(e_1)|\dot{E}(t)\cdot\vec{r}_1|\varepsilon_1(e_1)\rangle\langle\phi_2(e_2)|\varepsilon_2(e_2)\rangle + \langle\phi_1(e_2)|\dot{E}(t)\cdot\vec{r}_2|\varepsilon_1(e_2)\rangle\langle\phi_2(e_1)|\varepsilon_2(e_1)\rangle, \quad (1)$$

and the observed signal will contain cross terms between the two parts. The cross terms would be maximized when the two parts are of comparable magnitude, which could occur in the vicinity of the conical intersection when strong field coupling leads to charge separation.

Ab initio modeling of strongly nonadiabatic systems in strong fields is currently state of the art. We hope that the experimental observations reported here will be important in testing future models of strong-field ionization of photoexcited polyatomic molecules. The dominant processes in the interaction between atoms and strong laser fields are explained in a framework that uses a single active electron [30-32]. This is the case for non-sequential double ionization as well, where the first emitted electron accumulates enough kinetic energy in the ponderomotive potential of the strong laser field to eject a second electron during a rescattering [33]. Such rescattering may play a role in the explanation of the double ionization enhancement described here, but we cannot rule out a process that requires a breakdown of the single electron approximation. Previously reported quantum resonance rings [28], which arise from strong-field dressing of the crossing states in the vicinity of conical intersections, potentially play an additional role. Future experiments that benefit from higher time resolution and variation of the strong-field laser parameters will help understand the mechanism of the enhanced strong-field multiple ionization in the vicinity of conical intersections.

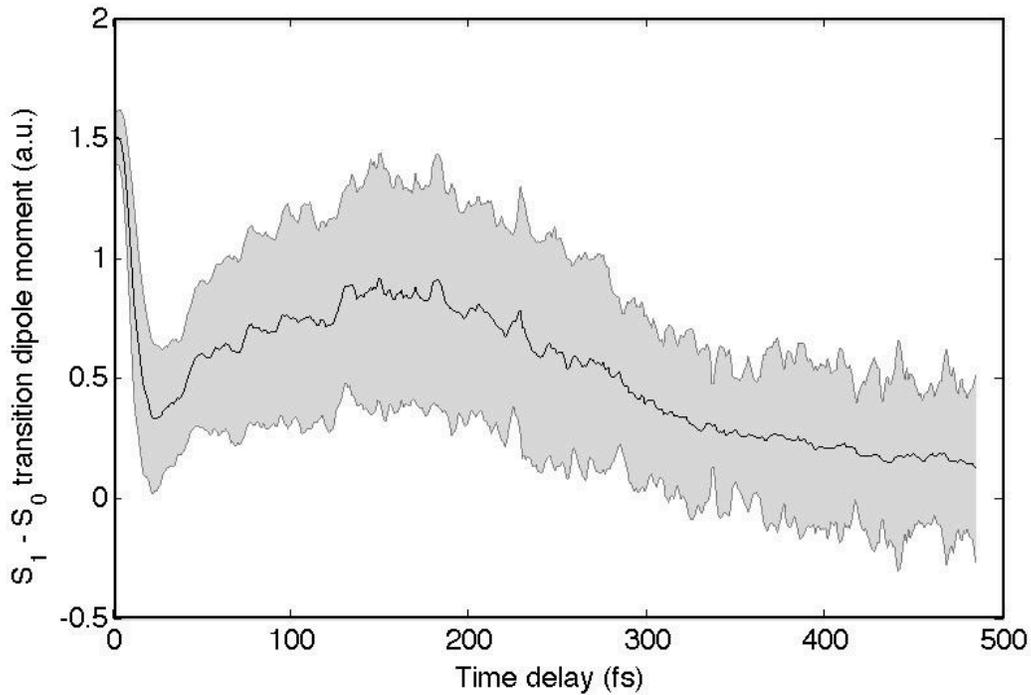

Fig. S1 Transition dipole moment between $S_1$ and $S_0$ states as a function of time delay upon the UV excitation, calculated using Ab Initio Multiple Spawning technique (see text for more details).

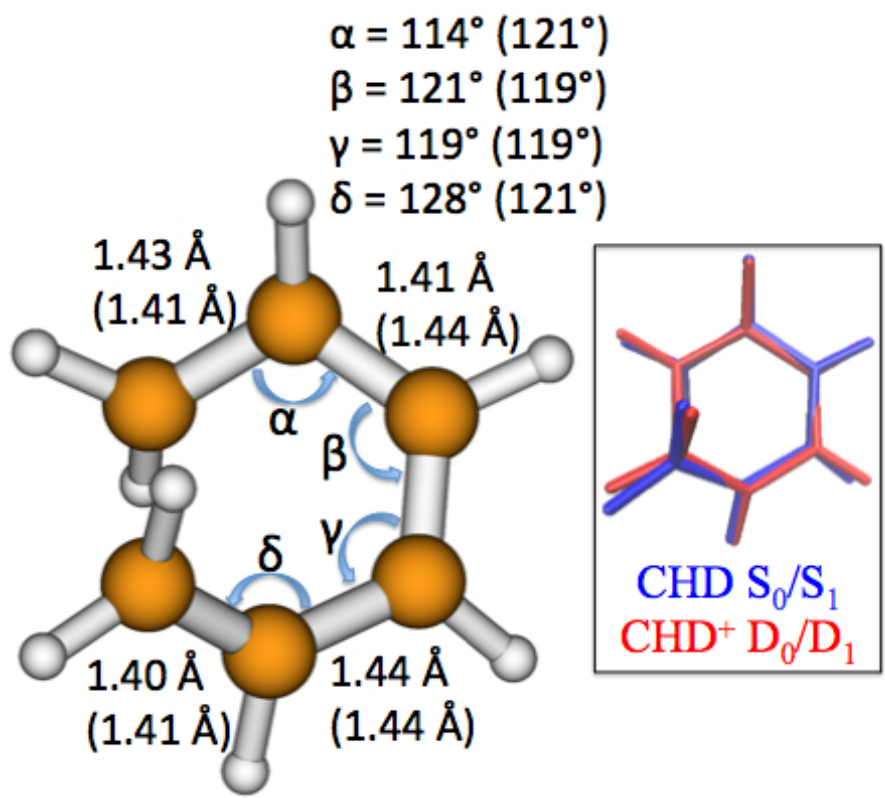

Fig. S2 The geometrical parameters of neutral and cation CHD at minimum energy conical intersection for $S_1/S_0$ and $D_1/D_0$, respectively. The cation parameters are shown in parentheses. The neutral and cation calculations were carried out with SA-2-CAS(6/6)-MSPT2/6-31G* and SA-2-CAS(5/6)-MSPT2/6-31G* methods, respectively. The inset shows a comparison of the geometries of the neutral (blue) and cation (red) CHD.